\documentclass[prb,reprint,showpacs,twocolumn,superscriptaddress,floatfix]{revtex4-1}
\usepackage{epsfig}
\usepackage{amsmath}
\usepackage{amssymb}
\usepackage{amsfonts}
\usepackage{mathptmx}
\usepackage{dcolumn}
\usepackage{eucal}
\usepackage{bm}
\usepackage{color}
\usepackage[colorlinks,linkcolor=blue,citecolor=blue]{hyperref}
\usepackage{epstopdf}
\usepackage{url}

\newcommand{\asym}{\mathcal{R}}

\begin{document}


\title{Microwave quantum refrigeration based on the Josephson effect}

\author{Paolo Solinas}
\affiliation{SPIN-CNR, Via Dodecaneso 33, 16146 Genova, Italy}
\author{Riccardo Bosisio}
\affiliation{SPIN-CNR, Via Dodecaneso 33, 16146 Genova, Italy}
\affiliation{NEST, Instituto Nanoscienze-CNR and Scuola Normale Superiore, I-56127 Pisa, Italy}
\author{Francesco Giazotto}
\affiliation{NEST, Instituto Nanoscienze-CNR and Scuola Normale Superiore, I-56127 Pisa, Italy}

\date{\today}

\begin{abstract}
We present a microwave quantum refrigeration principle based on the Josephson effect.
When a superconducting quantum interference device (SQUID) is pierced by a time-dependent magnetic flux, it induces changes in the macroscopic quantum phase and an effective finite bias voltage appears across the SQUID. This voltage can be used to actively cool well below the lattice temperature one of the superconducting electrodes forming the interferometer.
The achievable cooling performance combined with the simplicity and scalability intrinsic to the structure pave the way to a number of applications in quantum technology.
\end{abstract}

\maketitle

\section{ Introduction}
One of the key lessons we learn from themodynamics is that in order to extract heat from a system we must spend energy in the form of work.
We can thereby use this effect to cool that system.
However, as soon we generate a thermal gradient between the system and its surrounding, a heat flow opposite to the thermal gradient tends to restore thermodynamic equilibrium.
For these reasons, to be of practical use, we must be able to sustain over time the thermal gradient by performing continuously work on the system.
The simplest way to do this is to cyclicly drive the system out-of-equilibrium.
These principles are at the basis of any (macroscopic or microscopic) thermal machine and refrigerator.

In the push towards device miniaturization and quantum technologies, it has become of pivotal importance  the realization of high-performance nanoscale electronic  coolers \cite{Giazotto2006RMP,Muhonen2012}.
There are several successful solid-state quantum refrigeration schemes exploiting superconductors most of which are based either on normal metal-insulator-superconductor (NIS) \cite{Nahum,Leivo,Pekola2004,saira,Chauduri,Oneil,Nguyen13,Nguyen14,Lowell} or superconductor-insulator-superconductor (SIS) \cite{Manninen,Tirelli,Quaranta,Camarasa} tunnel junctions, even in combination with magnetic elements \cite{Giazotto2002,Giazotto2006,Ozaeta,Kawabata}.
In such systems, electronic refrigeration occurs thanks to the presence of the energy gap in the superconducting density of states. The latter provides an effective \emph{energy-filtering} mechanism yielding a substantial electron \emph{cooling} upon voltage biasing the tunnel junction near the gap edge \cite{Giazotto2006RMP,Muhonen2012}. 

\begin{figure}[t!]
    \begin{center}
    \includegraphics[scale=.38]{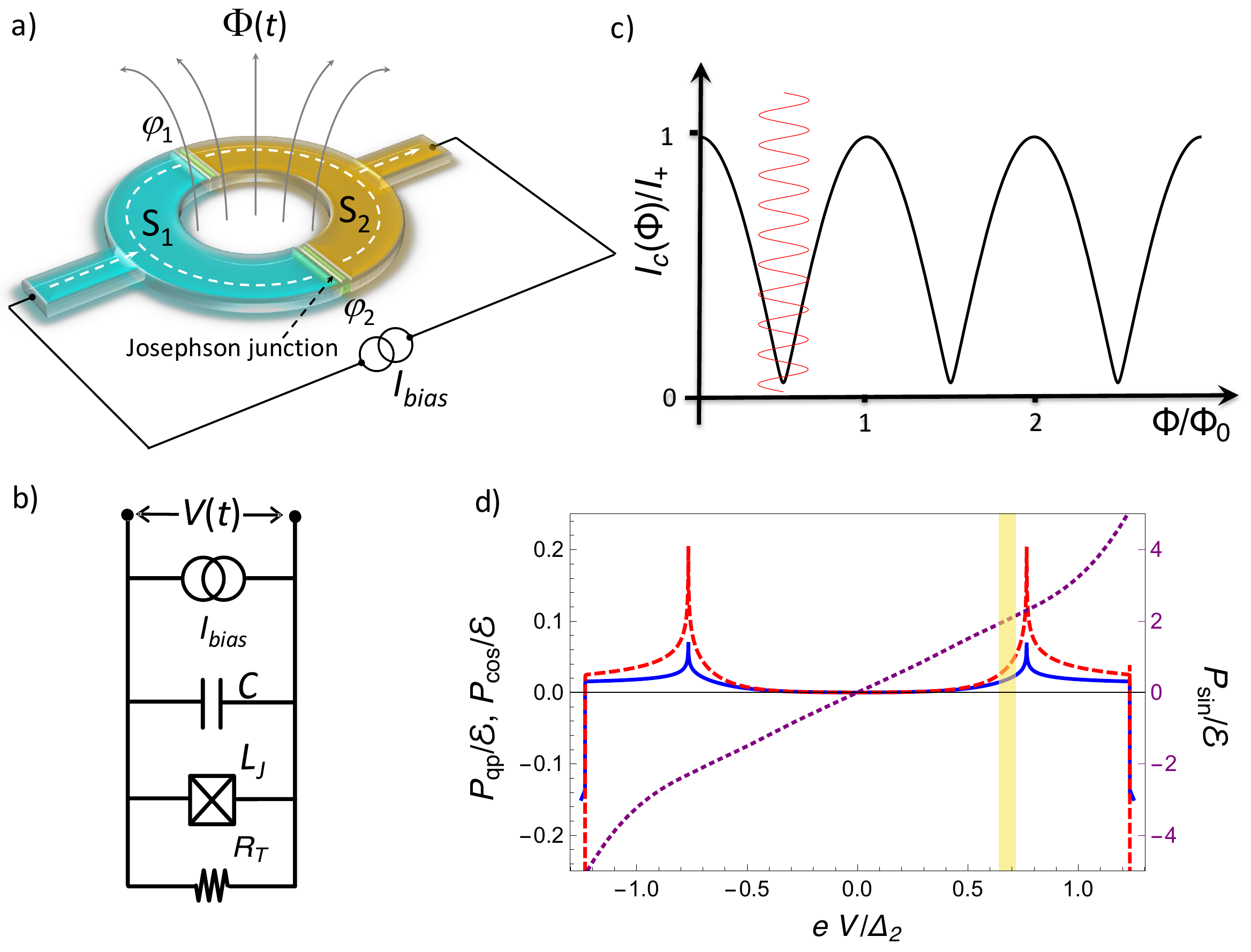}
   \end{center}
    \caption{ a) A superconducting quantum interference device (SQUID) pierced by a time-dependent magnetic flux $\Phi(t)$. 
$S_i$, $\Delta_i$, $T_i$, $\varphi_i$ denote the superconductor, the energy gap, the temperature and the superconducting phase difference, respectively.
$I_{bias}$ is the dissipationless supercurrent used to bias the interferometer.
     b) Equivalent electric circuital description of the SQUID. The parameter $C$, $R_T$, $L_J$ are the total capacitance, resistance and Josephson inductance of the SQUID, respectively; $I_{bias}$ is the biasing current and $V(t)$ is the effective voltage generated by the external drive.
     c) Critical current $I_C$ versus magnetic  flux $\Phi$ for an asymmetric SQUID (black line). $I_+$ is the maximum critical current of the SQUID, and $\Phi_0$ is the flux quantum. 
     The red curve represents the modulation of the magnetic flux centered around one of the interference nodes.
     d) Heat current flowing through a Josephson junction vs bias voltage $V$ for $\Delta_2/\Delta_1=3.3$: $P_{qp}$, $P_{\cos}$ and $P_{\sin}$ are represented in solid blue, red dashed and purple dotted curves, respectively.
       The yellow shaded region in the figure denotes the working voltage interval in the discussed example.
       Here, $\mathcal{E}=\Delta_2^2/(e^2 R_{T_i})$, $\Delta_2 = 200~\mu$eV, and $R_{T_i}$ is the normal-state resistance of each Josephson junction.
    } 
    \label{fig:fig1}
\end{figure}

Here, we propose and analyze the concept for a microwave Josephson refrigerator (MJR) [see Fig. \ref{fig:fig1} a)].
The structure we envision is a superconducting quantum interference device (SQUID) which allows us to control the dynamics of the macroscopic quantum phase through an externally applied time-dependent microwave magnetic field.
The  operating principle of this refrigeration method is based on the recent discovery that a driven SQUID can generate intense voltage pulses \cite{Solinas2015JRC, Solinas2015Radiationcomb,Bosisio2015Parasiticeffects}.
These voltage pulses can be used to actively transfer heat from one superconductor to the other and, therefore, to  \emph{cool} one of them.
The whole process is fully {\it phase-coherent} since it  critically depends on the induced dynamics of the superconducting phase.

For a realistic structure, we obtain sizeable cooling performance.
The superconducting electronic temperature can be reduced from $70\%$ to $40\%$ depending on the fabrication parameters, and the temperature working regime.
In particular, the MJR behavior depends on the resistance and capacitance of the SQUID junctions as well as on the gap engineering of the two superconductors forming the interferometer.
Depending on the final purpose these can be, in principle, fine-tuned to optimize this refrigeration scheme.
Important advantages of this cooling structure stem from the simplicity of its design, from its scalability, and from the fact it can be operated at a distance.
As a matter of fact, the only requirement is an external microwave magnetic field.
These facts open the way to a number of possible applications.
Connected to other quantum devices it can be used to remotely cool down them.
Furthermore, arrays of parallel MJRs can be engineered to extract heat from large devices, and to quickly and efficiently cool down them.

The paper is organized as follows. In Sec. \ref{sec:heat_current} and \ref{sec:dynamics} we discuss the device heat transport properties and its dynamics, respectively.
In Sec. \ref{sec:cooling_performances}, by using a power balance equation, we estimate the device cooling performances.
Section \ref{sec:conclusions} contains our conclusions.

\section{ Heat current}
\label{sec:heat_current}
We consider a SQUID [as shown in Fig. \ref{fig:fig1} a)] composed by two different superconductors $S_1$ and $S_2$.
Its electric and thermal state is characterized by the two superconducting phases $\varphi_1$ and $\varphi_2$ across the Josephson junctions (JJs).
We neglect the inductance of the superconducting loop so that the phases are related by the flux quantization condition, $\varphi_1 - \varphi_2 + 2 \pi \Phi/\Phi_0 = 2 \pi n$, where $n$ is an integer, $\Phi$ is the applied magnetic flux through  the SQUID  and  $\Phi_0\simeq 2\times 10^{-15}$ Wb is  the flux quantum.
The SQUID is connected to a generator that supplies a small \emph{non-dissipative} bias current $I_{bias}$  [see Fig. \ref{fig:fig1} a) and -b)].
Its only purpose is to give a preferred direction for the dynamics of the phase \cite{Solinas2015JRC, Solinas2015Radiationcomb,Bosisio2015Parasiticeffects}. 
A part from this, there is no need for additional connections, and the device can therefore be isolated.

The coherent thermal transport properties of Josephson tunnel junctions have been studied both theoretically \cite{golubev2013heat,Guttman1997Phase,Guttman1998Interference,Maki1965Entropy,Zhao2003Phase,Zhao2004Heat} and experimentally \cite{Giazotto2012Josephson,Giazotto2012Phasecontrolled,Giazotto2013coherentdiffraction, Fornieri2016heattransport}.
We denote with $V$ the voltage drop across the JJs, with $\Delta_i$ and $T_i$ the energy gap and the temperature of the $i$-th superconductor, respectively.
The heat current $P_i(t)$ flowing between two tunnel-coupled superconductors $S_1$ and $S_2$ through the $i$-th JJ consists of three contributions \cite{golubev2013heat,Guttman1997Phase,Guttman1998Interference,Maki1965Entropy,Zhao2003Phase,Zhao2004Heat},
\begin{equation}
 P_i(t)= P_{qp,i} (V) + P_{\cos,i} (V) \cos \varphi_i(t)+ 
 P_{\sin,i} (V) \sin \varphi_i(t).
\label{eq:P1}
\end{equation}
The powers $P_{qp,i} (V)$, $P_{\cos,i} (V)$ and $P_{\sin,i} (V)$ are the quasi-particle, and the anomalous heat currents, respectively.
They read 
\begin{widetext}
\begin{eqnarray}
 P_{qp,i} (V) &=& \frac{1}{e^2 R_{T_i}} \int dE~ N_1(E-e V) N_2(E) (E- e V) [f_1(E- e V) - f_2(E)] \nonumber \\
 P_{\cos,i} (V) &=& - \frac{1}{e^2 R_{T_i}} \int dE ~N_1(E-e V) N_2(E) \frac{\Delta_1 \Delta_2}{E} [f_1(E- e V) - f_2(E)] \nonumber \\
 P_{\sin,i} (V) &=& \frac{e V}{2 \pi e^2 R_{T_i}} \int d\epsilon_1  \int d\epsilon_2 \frac{\Delta_1 \Delta_2}{E_2} 
 \Big[ \frac{1-f_1(E_1) - f_2(E_2)}{(E_1+E_2)^2-e^2 V^2 } +\frac{f_1(E_1) - f_2(E_2)}{(E_1-E_2)^2-e^2 V^2 } \Big].
 \label{eq:heat_powers}
\end{eqnarray}
\end{widetext}
Here, $f_j(E) = 1/(1+ e^{E/k_BT_j})$ is the Fermi distribution function, 
$ N_j(E) = \left|\Re {\rm e} \left[ \frac{E + i \gamma}{\sqrt{[E+i \gamma]^2 -\Delta_j^2}} \right]\right|$ is the  smeared BCS density of states, $\gamma=10^{-4}\Delta_2$ is the Dynes broadening parameter \cite{Dynes1978,Pekola2004}, and  $R_{T_i}$ is the  normal-state  resistance of each junction composing the SQUID.

By choosing superconductors with different energy gaps allows us to create a thermal asymmetry in the structure \cite{Martinez2013Efficient, Martinez2015Rectification, Fornieri2015Electronicheat}.
Its effect is captured by the asymmetry parameter $r=\Delta_2/\Delta_1$ which, in the MJR, has the purpose to improve and optimize the device performance. 

An example of the heat current contributions (\ref{eq:heat_powers}) vs bias voltage across the SQUID is shown in Fig. \ref{fig:fig1} d) for $r=3.3$ and $T_2=T_1$.
To optimize the heat transport we will focus in the region around $V=(\Delta_2-\Delta_1)/e$ where the quasi-particle and cosine are maximal.

Equations  (\ref{eq:heat_powers}) remain valid even in presence of a time-dependent voltage if the quasi-particles can be considered in a Fermi distribution, i.e., not out of equilibrium.
This assumption is at the basis of any calculation and experiment based on the Josephson effect and it is usually satisfied even in presence of fast oscillating voltage \cite{Tinkham1996, Barone1982}.
Since the physical process underlying the heat and the charge transport is the same \cite{Barone1982}, the above equations should have similar vast range of validity.

\section{SQUID dynamics}
\label{sec:dynamics}
The dynamics of a driven SQUID can be complex and must be solved numerically.
We rely on the {\em driven } resistively and capacitively shunted Josephson junction (RCSJ) equation \cite{Solinas2015JRC, Solinas2015Radiationcomb,Bosisio2015Parasiticeffects}. 
By introducing the phase $\varphi = (\varphi_1+\varphi_2)/2$, the Josephson current flowing through the SQUID $I_J= I_{c_1}  \sin \varphi_1 + I_{c_2}  \sin \varphi_2$ can be written as $I_J[\varphi;\phi(\tau)] = I_+ [\cos \phi \sin \varphi + \asym~ \sin \phi \cos \varphi]$.
Here, $\phi = \pi \Phi/\Phi_0$ is the normalized applied magnetic flux, $I_+ = I_{c_1} + I_{c_2}$, $\asym = (I_{c_1} - I_{c_2})/ (I_{c_1}+I_{c_2})=(R_{T_2}-R_{T_1})/ (R_{T_1}+R_{T_2})$ (assuming that $I_{c_i} \propto 1/R_{T_i}$) and $I_{ci}$ is  the critical current of the $i$-th junction.
We assume junctions with the same capacitance $C_i$ but different critical current and resistance.
We can write  the RCSJ equation as
\begin{equation}
 \frac{\hbar C}{2 e }  \ddot{{\varphi}} + \frac{\hbar }{2 e R_T } \dot{\varphi} +  \frac{\hbar \mathcal{R}}{2 e R_T } \dot{\phi} + I_J[\varphi;\phi(\tau)] = I_{bias}
 \label{eq:varphi_RCSJ}
\end{equation} 
where $C=C_1+ C_2=2 C_1$ and $R_T = R_{T_1} R_{T_2}/(R_{T_1}+R_{T_2})$ are the total SQUID capacitance and resistance, respectively (see Appendix \ref{app:dynamics}).

The solution of Eq. (\ref{eq:varphi_RCSJ}) combined with the flux quantization condition gives immediately the dynamics of $\varphi_i$ and, through the Josephson relation, the voltage generated across the junctions ($V$).
From these, we can directly calculate the time-dependent heat current transferred across the $i$-th junction $P_i$, and the total heat current flowing through the SQUID as $P = \sum_{i=1,2} P_i$ \cite{Martnez2013Fullybalanced}.
Since the system is driven and has, in general, a complex dynamics, the relevant quantity is the average power transferred within a time interval $t_0$.
This is obtained by calculating the heat transferred as $P_{av} (t)= (1/ t_0) \int_t^{t+t_0} dt P$.
In the following, we consider a simple monochromatic drive of the magnetic field, $ \Phi(t) = \Phi_M \cos \left( 2 \pi \nu t \right) + \Phi_m $,
where $\nu$ is the drive frequency, while $\Phi_M$ and $\Phi_m$ are the maximum and minimum magnetic flux, respectively. 
\begin{figure}
    \begin{center}
    \includegraphics[scale=.6]{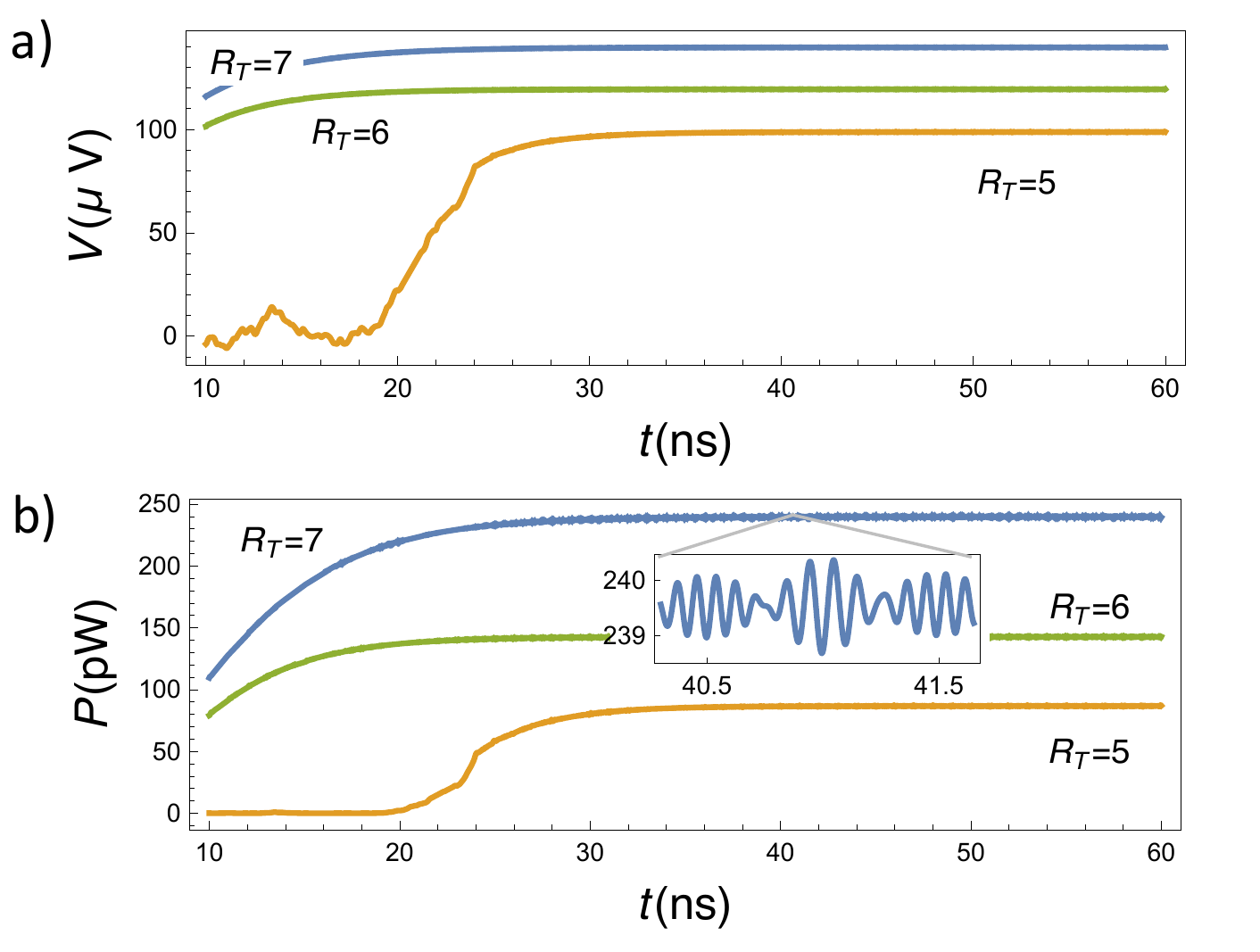}
   \end{center}
    \caption{ Time dependence of the voltage $V$ a) and of the heat current $P$ b) trough the SQUID for different tunneling resistances $R_T=7, 6, 5~$Ohm. 
    The parameters are $C=100~$pF, normalized bias current $I_{bias}/I_+=2\times10^{-2}$, and drive frequency $\nu=1~$GHz for a SQUID with junction asymmetry of $\asym=0.05$. The voltage dynamics is averaged over the heat transport time $\tau_{heat}=10~$ns.
    The inset in panel b) shows the oscillating behaviour of the heat current for $R_T=7~$Ohm.
    The resistances are expressed in Ohm.
    } 
    \label{fig:V_phase_time}
    \label{fig:fig2}
\end{figure} 

The dynamics of the phase $\varphi$ has very different behaviors depending on the parameters of the SQUID and the external drive.
For our purpose, they can be distinghuised by the presence or the absence of phase jumps \cite{Solinas2015JRC, Solinas2015Radiationcomb,Bosisio2015Parasiticeffects}.
We can select {\it a priori} which dynamics to induce by driving the magnetic flux across or avoiding an interference node of the critical current at $n \Phi_0/2$, as shown in Fig. \ref{fig:fig1} c) 
 \cite{Tinkham1996}.
In the \emph{absence} of crossing, the phase dynamics follows the drive modulation.
In such a case, from the Josephson relation, the effective voltage appearing across the SQUID is quite small (i.e., of the order of a fraction of $\mu V\sim 10^{-3}\Delta_2/e$ for $1~$GHz frequency drive).
For  a realistic drive source, it therefore leads to a somewhat limited heat transfer across the interferometer preventing an efficient electron cooling [see Fig. \ref{fig:fig1} d)].
The alternative choice is to let the magnetic flux to \emph{cross} an interference node, as displayed in Fig. \ref{fig:fig1} c).
Recently, it has been shown \cite{Solinas2015JRC,Solinas2015Radiationcomb,Bosisio2015Parasiticeffects} that when this occurs the superconducting phase undergoes a sequence of  fast $\pi$ jumps.
This corresponds to sharp voltage pulses developed across the SQUID junctions.
Under this condition, a moderate frequency drive can generate from some hundreds to thousands of higher harmonics \cite{Solinas2015JRC,Solinas2015Radiationcomb,Bosisio2015Parasiticeffects}.
This frequency up-conversion allows us to reach sufficiently high voltages and, thus, the peaks in Fig. \ref{fig:fig1} d) where the heat current and the cooling power are maximized.

The other important ingredient is to include a finite SQUID capacitance $C$ of the order of $\sim 50-100~$pF.
The latter introduces a time scale $\tau= R_TC$ in the RCSJ dynamics, and allows to sustain a large effective applied voltage across the interferometer over long times.
In other words, if $\tau$ is large enough, near voltage pulses are broadened until they merge together leading to a constant effective voltage applied across the SQUID.

The dynamics of the interferometer depends on the combination of the SQUID intrinsic parameters, i.e., $C$, $R_T$, $\mathcal{R}$ and $r$, and on the external ones, i.e., $\nu$.
Even if the details may change, we always observe that the system, after an initial transient, reaches a \emph{stationary} state characterized by constant voltage with fast oscillations superimposed [see Fig. \ref{fig:V_phase_time}].
These oscillations are usually fast with respect to the heat transport time $\tau_{heat}$ between superconductors.
This can be estimated of the order of $\sim10-10^3~$ns (see Ref. \cite{Rabani2008Phase} and Appendix \ref{app:entropy}).
Therefore the effective voltage relevant for thermal transport is averaged over this time-scale.
Examples of the voltage dynamics average over $\tau_{heat}=10~$ns are presented in Fig.  \ref{fig:V_phase_time} a) for different values of the tunneling resistance.
Notably, the stationary voltage seems to depends on $R_T$ and $C$ but not on the driving frequency as soon as $\nu >100~$MHz.
In our calculation we set a standard drive frequency of $\nu=1~$GHz.
Such a frequency allows us to safely neglect the heating due to the photon-assisted tunneling induced by the drive, as this usually becomes relevant above frequencies of the order of $10~$GHz \cite{Kopnin2008photon-assisted}.
The SQUID fabrication parameters can be tailored so that $V_{stat}$ is close to the matching peak at $(\Delta_2-\Delta_1)/e$ thereby maximizing the cooling power [see Fig. \ref{fig:fig1} d)]


An example of the instantaneous transferred power $P$ is plotted in Fig. \ref{fig:V_phase_time} b).
From this we can obtain the time averaged power $P_{av}$ in the stationary regime.
To calculate it we have taken different $t_0$ so to have $P_{av}$ independent of its specific value.
It is worthwhile to emphasize that in the stationary regime the phase grows linearly in time, i.e., $\varphi \propto V_{stat} t$.
Therefore, the average cosine and sine heat current terms \emph{vanish}, and thermal transport  essentially occurs thanks to the quasi-particle contribution.
Yet, we stress that this effect is completely \emph{phase coherent} since phase dynamics is the essential ingredient to enable heat transfer across the structure.

\section{Cooling performance}
\label{sec:cooling_performances}
Let us now analyze the cooling performance achievable in a realistic structure.
We consider the electrode $S_2$ to be large enough so that it can be treated as having infinite heat capacitance, and to be well thermalized with the lattice phonons  residing at bath temperature, $T_2=T_{bath}$.
Differently, the superconducting lead  $S_1$ is taken to be small so that any heat current flowing through it may easily change its temperature and cooled.
It is useful to first discuss the cooling process from the point of view of the SQUID alone.
If the two superconductors reside initially at the same temperature equal to the bath temperature, i.e., $T_1^{initial}=T_2=T_{bath}$, the driving magnetic field leads to quasiparticle cooling in $S_1$.
As soon as we establish a thermal gradient across the SQUID, the heat flows in the thermodynamical direction from the hotter to the colder superconductor.
The final stationary temperature and temperature gradient in the system is reached when these two competing effects balance each other.

In the above picture we have neglected the energy exchanged by quasiparticles in $S_1$ with the phononic bath.
The heat current flowing between the electrons at temperature $T_1$ and  lattice phonons  at temperature $T_{bath}$ is given by \cite{Timofeev2009,Maisi2013Excitation}
\begin{eqnarray}
 P_{qp-ph} (T_1,T_{bath}) = -\frac{ \Sigma \mathcal{V}}{ 96 \zeta(5) k_B^2} \int dE E  \int d\epsilon \epsilon^2 {\rm sgn} (\epsilon) L_{E,E+\epsilon} \nonumber \\
 \times \Big[ \coth \left( \frac{\epsilon }{2 k_B T_{bath}} \right) \left(f_E^{(1)} - f_{E+\epsilon}^{(1)}\right) - f_E^{(1)} f_{E+\epsilon}^{(1)} +1 \Big], \,\,\,\,\,\,\,\,\,\,\,\,\,\,\,\,\,\,
\label{eq:Pqp-ph}
\end{eqnarray}
where $f_E^{(1)} = f_1(-E) -f_1(E)$,  $L_{E,E'}= N(E) N(E') (1-\frac{\Delta_1^2}{E E'})$, $\Sigma=2\times10^{-8}~{\rm W}/({\rm m}^3~{\rm K}^5)$ is the electron-phonon coupling constant of Al \cite{Giazotto2006RMP}, and $\mathcal{V}=10^{-17}~{\rm m}^3$ is the $S_1$ electrode volume.
Furthermore, additional heating can come through the \emph{radiative} electron-photon heat exchange occurring between the two superconductors, $P_{\gamma} (T_1,T_2,\Phi)$ \cite{BosisioPhotonic2016}. The latter can be written as
\begin{equation}
P_{\gamma} (T_1,T_2,\Phi)=\int_0^{\infty}\frac{\text{d}\omega}{2\pi}\hbar \omega \mathcal{T}(\omega,T_1,T_2,\Phi)[n(\omega,T_2)-n(\omega,T_1)],
\end{equation}
where $n(\varepsilon,T)=[\text{exp}(\varepsilon /k_BT)-1]^{-1}$ is the Bose-Einstein photons distribution at temperature $T$, and $\mathcal{T}(\omega,T_1,T_2,\Phi)$ is the effective SQUID photonic transmission coefficient \cite{BosisioPhotonic2016}.
The final steady-state thermal balance equation which  must solved in order to obtain $T_1$ is therefore  $P_{av}(T_1,T_2)+ P_{qp-ph} (T_1,T_{bath})+P_{\gamma}^{av} (T_1,T_2)=0$ \cite{Giazotto2006RMP}.
Here, $P_{\gamma}^{av} (T_1,T_2)=(1/t_0)\int_t^{t+t_0}dt P_{\gamma}(T_1,T_2,\Phi)$ is the average transferred radiative power within a time interval $t_0$.
Its time-dependency comes from the modulation of the magnetic flux $\Phi$. Since  $\Phi$ is periodically driven close to $n \Phi_0/2$, $P_{\gamma}\approx  P_{\gamma}^{av} $.

\begin{figure}
    \begin{center}
    \includegraphics[scale=.55]{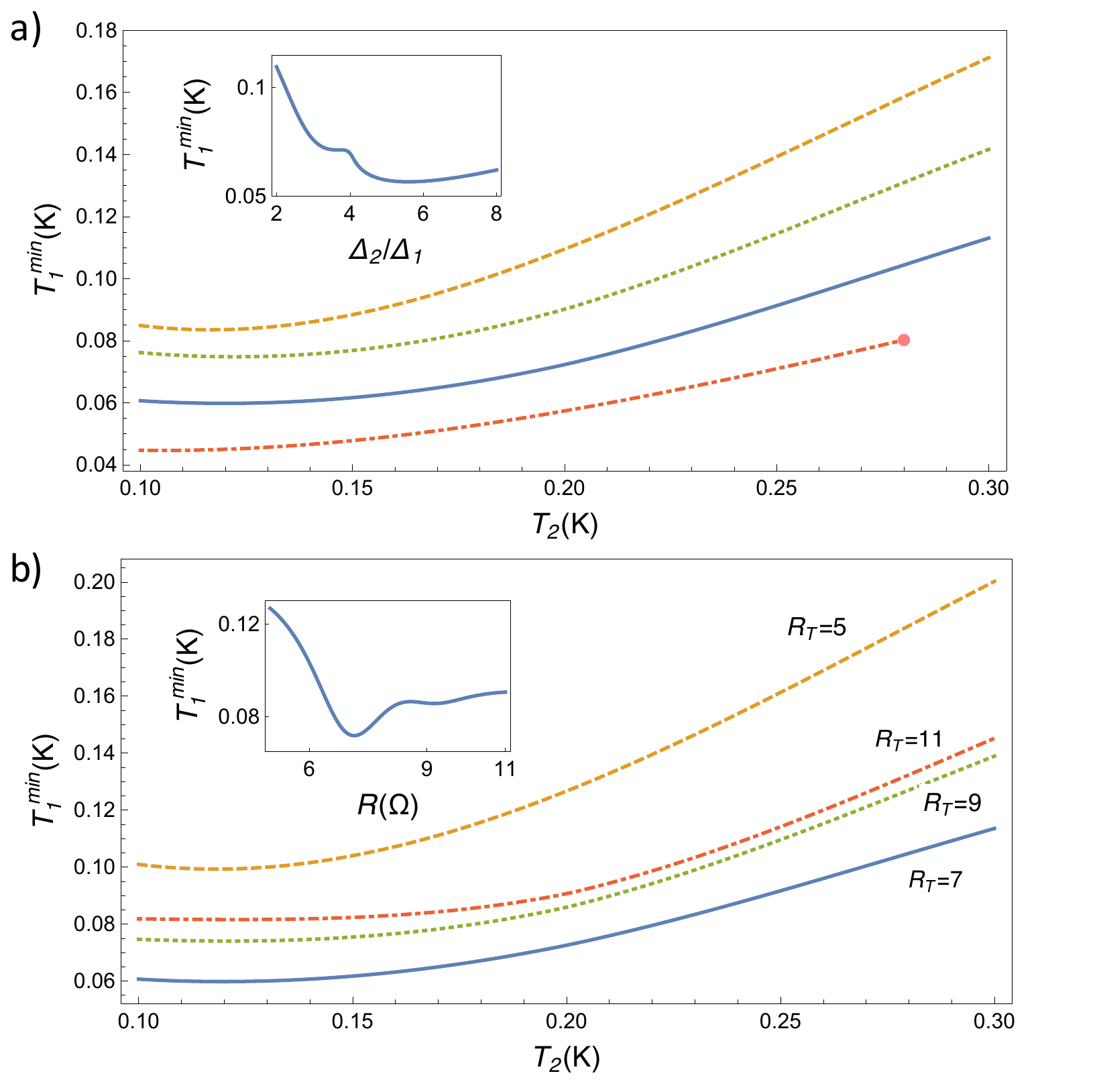}
   \end{center}
    \caption{a) Minimum achievable electron temperature $T_1^{min}$ vs $T_2=T_{bath}$. 
    The curves refer to different superconducting gap ratios: $\Delta_2/\Delta_1=2, 2.5, 3.3$ and $5$ from top to bottom.
    The red dot in the $\Delta_2/\Delta_1=5$ curve represents the temperature at which $S_1$ becomes a normal metal.
    Inset: minimum temperature $T_1^{min}$ reached as a function of $\Delta_2/\Delta_1$ at $T_2=200~$mK.
     b) Minimum achievable electron  temperature $T_1^{min}$ as a function of $T_2$ for different junction resistance $R_T$ values (expressed in Ohm).
    Inset: minimum temperature $T_1^{min}$ reached as a function of $R_T$ at $T_2=200~$mK.
    For these calculations we set $\Delta_2/\Delta_1=3.3$.
    } 
    \label{fig:fig3}
\end{figure}

The performances of the MJR are shown in Fig. \ref{fig:fig3}.
In particular,  Fig. \ref{fig:fig3} a)  displays the final temperature $T_1^{min}$ as a function of $T_2=T_{bath}=T_1^i$ for different $\Delta_2/\Delta_1$ ratios.
As a prototypical refrigerator we choose the one with $\Delta_2/\Delta_1=3.3$.
In this case, the cooling performance ranges from $\Delta T_1=T_1^i-T_1^{min} =186~$mk at $T_2=300~$mK to $\Delta T_1 =40~$mk at $T_2=100~$mK.
In general, the cooling process is more efficient at higher $T_2$, and for strongly asymmetric superconductors, i.e., for large $\Delta_2/\Delta_1$ ratios.
However, it seems that for $\Delta_2/\Delta_1> 5$ the achievable minimum temperature
  tends to saturate and no substantial improvement in cooling is obtained [see the inset of Fig. \ref{fig:fig3} a)].

The final achievable minimum temperature $T_1^{min}$ as a function of $T_2=T_{bath}=T_1^i$ and for different $R_T$ values is shown in Fig. \ref{fig:fig3} b).
Here we notice a non-monotonic behaviour as a function of $R_T$ [see also the inset in Fig. \ref{fig:fig3} b)].
This can be explained on the basis of the heat current [see Fig. \ref{fig:fig1} d)] and the voltage dynamics [see Fig. \ref{fig:fig2} a)].
Keeping fixed all the other parameters, the stationary voltage increases with the resistance $R_T$  [see Fig. \ref{fig:fig2} a)].
Therefore, an increase in junction resistance allows us to reach the maximum cooling power [at $(\Delta_2-\Delta_1)/e$]  for $R_T=7~$Ohm [as shown in Fig. \ref{fig:fig2} b)].
From this value any further resistance enhancement would lead to a decrease of the heat current, and to a worsening of the cooling performance.
This feature is well-captured by the plot shown in the inset of Fig. \ref{fig:fig3} b).

The performance of the present cooling principle can be compared to that of other time-dependent refrigeration methods.
For instance, in Ref. \cite{Pekola2007} with a Coulombic single-electron refrigerator (SER), 
 $\Delta T_1 \sim130~$mk at $T_2=300~$mK and $\Delta T_1 \sim30~$mk at $T_2=100~$mK were in principle achievable. Therefore, an optimized  MJR may outperform the SER in all the considered temperature ranges.
In addition, the  MJR has other practical advantages.
First, the structure stands out for the simplicity of fabrication and control.
Second, the superconductor can be cooled at a distance.
Third, due to its scalability and flexibility, it can be assembled to respond to different needs.
For instance, one can envision a network of parallel MJRs yielding a large cooling power.
Yet, the exploitation of the MJR depends on the temperature one intends to achieve.
It can be used as direct electron cooler if we are planning to work at temperatures around $100$ or $50~$mK.
Alternatively, it can be used as an efficient intermediate-stage electron refrigerator.

\section {Conclusions}
\label{sec:conclusions}
In summary, we have proposed a principle of coherent electron cooling based on the Josephson effect.
The microwave Josephson refrigerator is build from a SQUID made of superconductors with different gaps,
and  exploits the work performed by a microwave magnetic field to efficiently cool the superconductor with smaller energy gap.
The working principle stems from the dynamics induced in the macroscopic quantum phase by an external time-dependent magnetic drive. 
The latter yields a finite effective voltage drop appearing across the SQUID which enables electron cooling of one of the superconductors.
Finally, the MJR can be tuned by fabrication to reach  optimal cooling performance.

\begin{acknowledgments}
Fruitful discussions with C. Altimiras, S. Gasparinetti and A. Fornieri are gratefully acknowledged.
P.S. and R.B. have received funding from the European Union FP7/2007-2013 under REA
grant agreement no 630925 -- COHEAT and from MIUR-FIRB2013 -- Project Coca (Grant
No.~RBFR1379UX). 
F.G. acknowledges the European Research Council under the European Union's Seventh Framework Program (FP7/2007-2013)/ERC Grant agreement No.~615187-COMANCHE for partial financial support.
\end{acknowledgments}


\appendix

\section{Dynamics of the asymmetric SQUID}
\label{app:dynamics}

To describe the dynamics of the driven SQUID we rely on the resistively and capacitively shunted Josephson junction (RCSJ) equation.
The current $I_i$ flowing through the $i-$th junction is \cite{Tinkham1996}
\begin{equation}
  I_i=\frac{\hbar C_i}{2 e }  \ddot{{\varphi}}_i + \frac{\hbar}{2 e R_{T_i}} \dot{\varphi}_i + I_{J_i}[\varphi_i;\phi_i(\tau)],
  \label{app_eq:varphi_RCSJ_single_junc}
\end{equation}
where $C_i$, $ R_{T_i}$, $I_{J_i} = I_{c_i}  \sin \varphi_i$ and $\varphi_i$ are the capacitance, resistance, Josephson current and superconducting phase across the junction, respectively.  The parameter $I_{c_i}$ is  the critical current of the $i$-th junction.

The phases $\varphi_i$ are related through the flux quantization condition $\varphi_1 - \varphi_2 + 2 \pi \Phi/\Phi_0 = 2 \pi n$, where $n$ is an integer, $\Phi$ is the applied magnetic flux through  the SQUID  and  $\Phi_0\simeq 2\times 10^{-15}$ Wb is  the flux quantum.
By introducing the phase $\varphi = (\varphi_1+\varphi_2)/2$ and the normalized applied magnetic flux $\phi = \pi \Phi/\Phi_0$, we have that $\varphi_1 = \varphi + \phi$ and $\varphi_2 = \varphi - \phi$.

If the SQUID is biased with a current $I_{bias}$, the total current passing though the interferometer, i.e., $I_1+I_2=I_{bias}$, can be written as 
\begin{widetext}
\begin{equation}
 \frac{\hbar (C_1+C_2)}{2 e }  \ddot{{\varphi}} + \frac{\hbar (C_1-C_2)}{2 e }  \ddot{{\phi}} + \frac{\hbar (R_{T_1}+R_{T_2})}{2 e R_{T_1} R_{T_2} } \dot{\varphi} -  \frac{\hbar (R_{T_1}-R_{T_2})}{2 e R_{T_1} R_{T_2} } \dot{\phi} +  (I_{c_1} + I_{c_2}) \cos \phi \sin \varphi   +  (I_{c_1} - I_{c_2}) \sin \phi \cos \varphi = I_{bias}
  \label{app_eq:varphi_RCSJ_general}
\end{equation} 
\end{widetext}
The asymmetry in the SQUID is captured by the factor $\asym = (I_{c_1} - I_{c_2})/ (I_{c_1}+I_{c_2}) =(R_{T_2}-R_{T_1})/ (R_{T_1}+R_{T_2})$ (assuming that $I_{c_i} \propto 1/R_{T_i}$).
Assuming that $C_1=C_2$ and introducing the parameter $I_+ = I_{c_1} + I_{c_2}$, $C=C_1+C_2= 2 C_1$, $R_T = R_{T_1} R_{T_2}/(R_{T_1}+R_{T_2})$, we have 
\begin{equation}
 \frac{\hbar C}{2 e }  \ddot{{\varphi}} + \frac{\hbar }{2 e R_T } \dot{\varphi} +  \frac{\hbar \mathcal{R}}{2 e R_T } \dot{\phi} +  I_+ \left( \cos \phi \sin \varphi   + \mathcal{R} \sin \phi \cos \varphi \right) = I_{bias}.
 \label{app_eq:varphi_RCSJ_general}
\end{equation} 

\section{Heat transfer time-scale}
\label{app:entropy}

For a driven system the voltage applied to the device can be obtained by Eq. (\ref{app_eq:varphi_RCSJ_general}).
We consider a simple monochromatic drive of the magnetic field, $ \Phi(t) = \Phi_M \cos \left( 2 \pi \nu t \right) + \Phi_m $, where $\nu=1~$GHz is the drive frequency, while $\Phi_M$ and $\Phi_m$ are the maximum and minimum magnetic flux, respectively. 

The drive is taken to cross the critical current interference node at $\Phi_0/2$.
Under these conditions, the phase dynamics can be complex as discussed in Refs. \cite{Solinas2015JRC, Solinas2015Radiationcomb,Bosisio2015Parasiticeffects} and higher harmonics of the fundamental frequency can be generated.
For this reason, the voltage across the SQUID can show modulation over short time-scales (below $1~$ns).

However, the interferometer cannot react and transport heat over such time scale and, thus, the effective voltage for thermal transport in Eqs. (\ref{app_eq:varphi_RCSJ_general}) in the main text is averaged over a heat transport time-scale $\tau_{heat}$.
This thermal time-scale can be estimated in the following way.
The \emph{electronic} entropy of the small superconductor $S_1$ is given by \cite{Rabani2008Phase}
\begin{eqnarray}
 S &=& -4 k_B N_F \mathcal{V} \int_0^{\infty} d \epsilon N(E) \Big[ (1- f(\epsilon, T_1)) \log(1- f(\epsilon, T_1)) \nonumber \\
 &&+(f(\epsilon, T_1) \log(f(\epsilon, T_1))
 \Big],
\end{eqnarray}
where $k_B$ is the Boltzmann constant, $ N_F$ is the density of states at the Fermi energy, $\mathcal{V}$  and $ \Delta_1$ are the volume and the energy gap of the superconducting electrode $S_1$. 
The function $f(\epsilon, T_1)$ is the Fermi-Dirac energy distribution,  $N(E) = \left|\Re {\rm e} \left[ \frac{E + i \gamma}{\sqrt{[E+i \gamma]^2 -\Delta_j^2}} \right]\right|$ is the  smeared BCS density of states, and $\gamma$ is the Dynes broadening parameter.

At temperature $T_1$, the heat transferred is $Q= S T_1$.
By supposing a constant quasi-particle heat current $P_{qp}$, we have that $Q= P_{qp} \tau_{heat}$.
To provide an estimate for $\tau_{heat}$ we can take $S \propto 4 k_B N_F \mathcal{V} \Delta_1$ and $P_{qp} \propto \Delta_1^2/(e^2 R_T)$, so that we obtain
\begin{equation}
 \tau_{heat} = \frac{4 k_B N_F \mathcal{V} e^2 R_T}{\Delta_1} T_1.
\end{equation}
By taking $N_F = 10^{47} J^{-1}m^{-3}$,  $\mathcal{V}= 10^{-17}$m$^3$, $R_T=1-10~$Ohm, $\Delta_1\approx 3.4-0.6 ~\times 10^{-23}~J$ and $T_1=100~$K, we get $\tau_{heat} \approx 10-10^3~$ns.
For the numerical analysis we have chosen the lowest value of $\tau_{heat}=10~$ns.


\bibliographystyle{apsrev4-1}
%

\end{document}